\documentclass[structabstract]{aa}  
\usepackage{graphicx}
\usepackage{txfonts}
\usepackage{natbib}
\usepackage{longtable}
\def\pdeg{\ifmmode $\setbox0=\hbox{$^{\circ}$}\rlap{\hskip.11\wd0 .}$^{\circ}
  \else \setbox0=\hbox{$^{\circ}$}\rlap{\hskip.11\wd0 .}$^{\circ}$\fi}

\begin{document}
\title{Dynamical cloud formation traced by atomic and molecular gas}


   \author{H.~Beuther
          \inst{1}
          \and
          Y.~Wang
          \inst{1}
          \and
          J.D.~Soler
          \inst{1}
           \and
          H.~Linz
          \inst{1}
           \and
          J.~Henshaw
          \inst{1}
          \and
          E.~Vazquez-Semadeni
          \inst{2}
          \and
          G.~Gomez
          \inst{2}
          \and
          S.~Ragan
          \inst{3}
          \and
         Th. Henning
          \inst{1}
           \and
          S.C.O.~Glover
          \inst{4}
           \and
          M.-Y.~Lee
          \inst{5}
           \and
          R.~G\"usten
          \inst{6}
}
   \institute{$^1$ Max Planck Institute for Astronomy, K\"onigstuhl 17,
     69117 Heidelberg, Germany, \email{name@mpia.de}\\
              $^2$ Instituto de Radioastronomía y Astrofísica, Universidad Nacional Autonoma de Mexico, Morelia, Michoacan 58089, Mexico\\
              $^3$ School of Physics and Astronomy, Cardiff University, Queen's Buildings, The Parade, Cardiff CF24 3AA, UK\\
              $^4$ University of Heidelberg, Institute for Theoretical Astrophysics, Albert-Ueberle-Str.~2, 69120 Heidelberg, Germany\\
              $^5$ Korea Astronomy and Space Science Institute, 776 Daedeok-daero, 34055 Daejeon, Republic of Korea\\
              $^6$ Max-Planck-Institut f\"ur Radioastronomie, Auf dem H\"ugel 69, 53121 Bonn, Germany
}

   \date{Version of \today}

\abstract
{Atomic and molecular cloud formation is a dynamical process. However,
  kinematic signatures of these processes are still observationally
  poorly constrained.}
{Identify and characterize the cloud formation signatures in atomic
  and molecular gas.}
{Targeting the cloud-scale environment of the prototypical infrared dark
  cloud G28.3, we employ spectral line imaging observations of the two
  atomic lines HI and [CI] as well as molecular lines observations in
  $^{13}$CO in the 1--0 and 3--2 transitions. The analysis comprises
  investigations of the kinematic properties of the different tracers,
  estimates of the mass flow rates, velocity structure functions, a
  Histogram of Oriented Gradients (HOG) study as well as comparisons to
  simulations.}
{The central IRDC is embedded in a more diffuse envelope of cold
  neutral medium (CNM) traced by HI self-absorption (HISA) and
  molecular gas. The spectral line data as well as the HOG and
  structure function analysis indicate a possible kinematic decoupling
  of the HI from the other gas compounds. Spectral analysis and
  position-velocity diagrams reveal two velocity components that
  converge at the position of the IRDC. Estimated mass flow rates
  appear rather constant from the cloud edge toward the center. The
  velocity structure function analysis is consistent with gas flows
  being dominated by the formation of hierarchical structures.}
{The observations and analysis are consistent with a picture where the
  IRDC G28 is formed at the center of two converging gas flows. While
  the approximately constant mass flow rates are consistent with a
  self-similar, gravitationally driven collapse of the cloud, external
  compression by, e.g., spiral arm shocks or supernovae explosions
  cannot be excluded yet. Future investigations should aim at
  differentiating the origin of such converging gas flows.}
\titlerunning{Cloud formation signatures}

\maketitle
  
\section{Introduction}
\label{intro}

Molecular clouds are formed out of the atomic phase of the
interstellar medium (ISM). While models detailing the transition from
the atomic to molecular phase exist (e.g.,
\citealt{franco1986,hartmann2001,bergin2004,krumholz2008c,krumholz2009b,sternberg2014,bialy2017b}),
the observational constraints of the molecular cloud formation
processes are still not properly characterized. Of particular
importance is the dynamic state of the clouds, i.e., whether they are
dominated by converging gas flows that could create over-densities
important for the atomic-to-molecular gas conversion (e.g.,
\citealt{koyama2000,audit2005,audit2010,vazquez2011,vazquez2019,
  heitsch2005,heitsch2006,heitsch2008b,banerjee2009,clark2012,gomez2014,motte2014,heiner2015,henshaw2016b,langer2017}),
or whether more quasi-static cloud contraction processes take place
where ever increasing densities may be related to the phase
transitions between the atomic and molecular gas (e.g.,
\citealt{elmegreen1993b,williams2000,mckee2007}).

\begin{figure}[htb]
\includegraphics[width=0.49\textwidth]{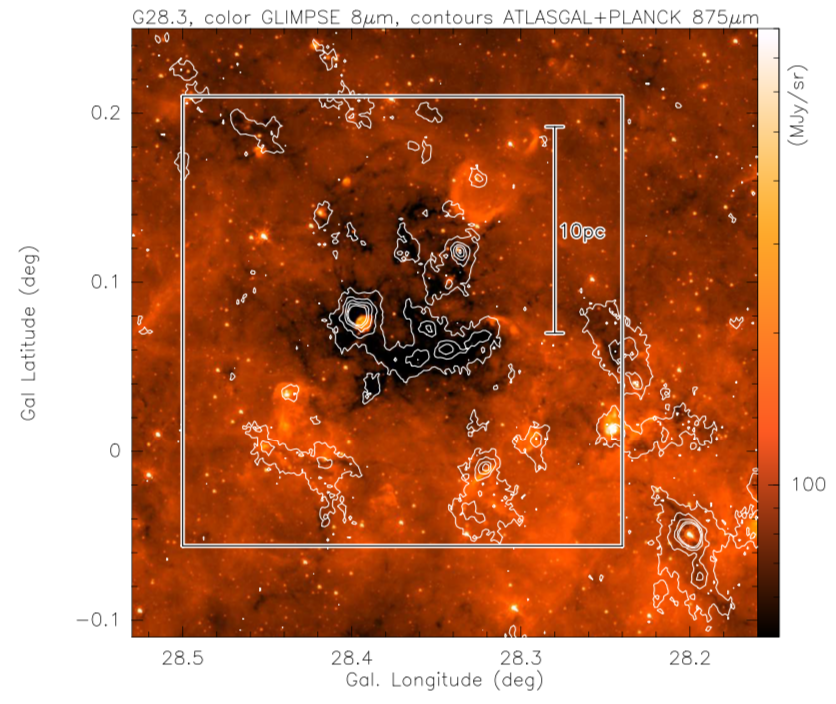}
\caption{Overview of the target region G28.3. The color-scale presents
  the 8\,$\mu$m emission from GLIMPSE \citep{churchwell2009}, and the
  contours show the 875\,$\mu$m emission from the ATLASGAL survey
  (\citealt{schuller2009}, contour levels start at $4\sigma$ level of
  200\,mJy\,beam$^{-1}$ and continue in $8\sigma$ steps up to
  1.8\,Jy\,beam$^{-1}$). The white box outlines the approximate region
  imaged with APEX in the [CI] and $^{13}$CO(3--2) emission. A scale
  bar is shown within the box.}
\label{8mu} 
\end{figure} 

In dynamical pictures of converging gas flows, cloud-cloud collisions
and cloud collapse flows, a convergence of gas flows towards some
point in space is involved. A major difference between converging gas
flows and cloud-cloud collisions on the one side, and cloud collapse
on the other side is that in the converging gas flow and cloud-cloud
collision picture the compression is produced by some external cause
(e.g., supernovae or spiral arm potentials, e.g.,
\citealt{maclow2004,haworth2018,kobayashi2018}) whereas the collapsing
cloud flows are dominated by the self-gravity of the cloud itself
(e.g., \citealt{vazquez2019}). In reality, externally driven
converging flows may produce the clouds that then further collapse
under their own self-gravity.

The observational characterization of the transition from atomic to
molecular gas requires observations of both phases in atomic and
molecular spectral lines, respectively.  Most molecular cloud studies
have been conducted in different lines of carbon monoxide covering a
large range of spatial scales (e.g.,
\citealt{dame2001,jackson2006,dempsey2013,barnes2015,heyer2015,rigby2016,umemoto2017,schuller2017}).
Similarly, the HI 21\,cm line has been observed a lot since its first
detection \citep{ewen1951}, notable surveys of the Milky Way are
\citet{kalberla2005,stil2006,mcclure-griffiths2009,kerp2011,winkel2016,hi4pi2016,beuther2016,wang2020a}.
However, kinematic studies of the HI emission are difficult because
the spectra typically cover velocity ranges of 100\,km\,s$^{-1}$ and
more. With such broad line emission that traces a mixture of the cold
neutral medium (CNM) and the warm neutral medium (WNM), unambiguous
identification of velocity structure belonging to individual clouds is
challenging (e.g.,
\citealt{kalberla2009,winkel2016,beuther2016,murray2018,wang2020a}). A
different approach to study the atomic hydrogen kinematics is to look
for HI self-absorption (HISA) features at comparably high HI column
densities against HI background emission. Such HISA features allow us
to isolate the CNM from the WNM. Some example studies in that
direction were conducted by, e.g.,
\citet{gibson2000,gibson2005,gibson2005b,li2003,krco2008,heiner2015,wang2020b}. In
this work, we will employ the HI/OH/Recombination line survey of the
Milky Way (THOR, \citealt{beuther2016,wang2020a}) to study the HI
kinematics from HISA features around the famous infrared dark cloud
(IRDC) G28.3 (Fig.~\ref{8mu}).

Neither molecular emission nor atomic HI emission trace the transition
phase between the two media well. While the ionized carbon [CII] may
trace that transition, [CII] emission depends on the local radiation
field strength. In infrared dark clouds like the target cloud of this
study, the [CII] emission is often too faint to be well detectable
(e.g., \citealt{beuther2014,clark2019}). Furthermore, since [CII]
cannot be observed from the ground, conducting large and sensitive
maps are often difficult to obtain. Therefore, one of the arguably
best tracers of the cloud formation and transition phase is the atomic
carbon fine structure line [CI] around 492\,GHz (e.g.,
\citealt{papadopoulos2004,offner2014,glover2015}). The [CI] emission
is believed to trace both the molecular clouds and also the more
diffuse emission from the CO-dark molecular gas as well as the atomic
envelope around the molecular clouds (e.g.,
\citealt{stoerzer1997,papadopoulos2004,offner2014,glover2015}).

Previous atomic carbon fine structure line studies typically covered
only comparably small areas not extending far into the more diffuse
cloud envelope structures (e.g.,
\citealt{keene1995,schilke1995,ossenkopf2011}). In an attempt to study
the early evolutionary stages of high-mass star formation, we
investigated four IRDCs in ionized, atomic and molecular carbon
\citep{beuther2014}. In one of the clouds (G48.66) we found evidence
for converging gas flows in [CII] emission. However, even in that
case, the size of the maps were comparably small not reaching far into
the environmental cloud.

Here we are presenting an atomic carbon [CI] study of the prototypical
IRDC G28.3 at scales larger of roughly $15'\times 15'$ or $\sim
20\times 20$\,pc$^2$. The molecular cloud G28.3 is a large region
where the innermost region is the well-known IRDC G28.3 (e.g.\,
\citealt{pillai2006,wang2008,ragan2012b,butler2012,butler2014,tan2013,tackenberg2014,kainulainen2013b,zhang2015,feng2016b}). Infrared
data from the Spitzer satellite show that filamentary extinction
structures extend from the large-scale atomic/molecular outskirts down
to the innermost cloud center (Fig.~\ref{8mu},
\citealt{churchwell2009}). At a kinematic distance of 4.7\,kpc,
observations of the dense gas in N$_2$H$^+$ indicate that the region
is in a stage of global collapse \citep{tackenberg2014}, and different
signs of star formation activity exist throughout the cloud (e.g.,
\citealt{wang2008,butler2012,tan2016,feng2016a}). The N$_2$H$^+$(1--0)
data of the central regions indicate already a velocity spread with
peak velocities varying approximately between 77.5 and
81.5\,km\,s$^{-1}$ \citep{tackenberg2014}. In the following, we use as
approximate velocity of rest $\varv_{\rm lsr}\sim
79.5$\,km\,s$^{-1}$. The above characteristics make the G28.3 complex
an ideal candidate to investigate the cloud formation and atomic to
molecular gas conversion processes.

In the following, we combine an analysis from the diffuse environmental
atomic and molecular cloud traced by HISA and $^{13}$CO(1--0) emission
to the denser G28.3 IRDC center better studied in the [CI] and
$^{13}$CO(3--2) transitions.

\section{Observations and data}
\label{data}

\subsection{THOR HI and $^{13}$CO(1--0) GRS data}

The HI data are taken from the HI/OH/Recombination lines survey of the
Milky Way (THOR, \citealt{beuther2016,wang2020a}) conducted with the
Very Large Array (VLA). The final atomic hydrogen HI data product is
the combined data cube from the THOR C-array observations with the
previous VLA D-array survey and GBT single-dish data (VGPS,
\citealt{stil2006}). The angular and spectral resolution as well as
the typical rms of the combined dataset are $40''$, 1.5\,km\,s$^{-1}$
and 10\,mJy\,beam$^{-1}$, respectively. For more details about the
survey and data products, see \citet{beuther2016}, \citet{wang2020a}
and the web-site at {\sf http://www.mpia.de/thor}.

The corresponding large-scale $^{13}$CO(1--0) data are taken from the
Galactic Ring Survey GRS, conducted with the FCRAO
\citep{jackson2006}. The spatial and spectral resolution as well as
the typical rms sensitivity of this survey are $46''$,
0.21\,km\,s$^{-1}$, and 0.13\,K, respectively.

\subsection{APEX [CI] and $^{13}$CO(3--2) observations}

\begin{figure}[htb]
\includegraphics[width=0.49\textwidth]{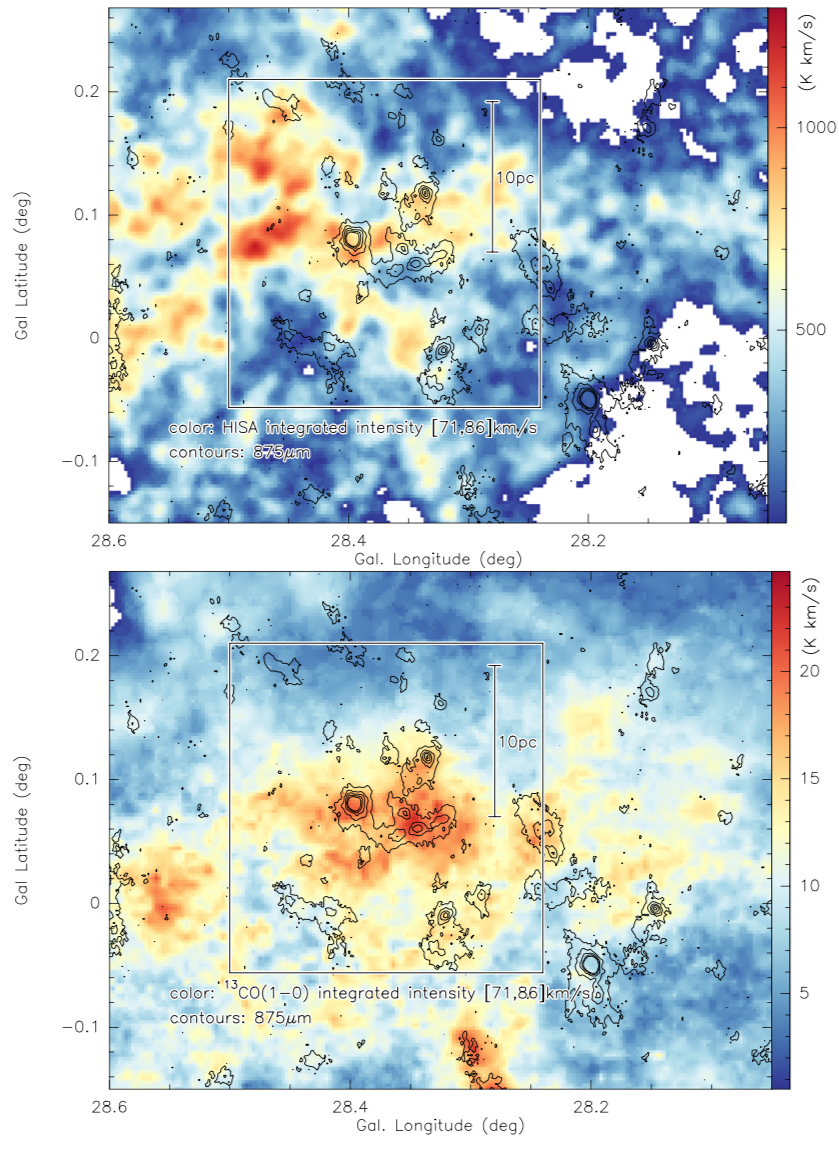}
\caption{Large-scale structure of the cold neutral medium (CNM)
  measured as HISA in the THOR survey (top panel,
  \citealt{beuther2016,wang2020a,wang2020b}) and the $^{13}$CO(1--0)
  emission from GRS (bottom panel, \citealt{jackson2006}). The
  color-scales show the inverted HISA emission and $^{13}$CO(1--0)
  maps, respectively (integrated from 71 to 86\,km\,s$^{-1}$ above
  15\,K in the HISA spectrum), and the contours show the 875\,$\mu$m
  emission from the ATLASGAL survey (\citealt{schuller2009}, contour
  levels start at $4\sigma$ level of 200\,mJy\,beam$^{-1}$ and
  continue in $8\sigma$ steps up to 1.8\,Jy\,beam$^{-1}$). The white
  box outlines the approximate region imaged with APEX in the [CI] and
  $^{13}$CO(3--2) emission. A scale bar is shown within the box.}
\label{hisa} 
\end{figure}

The atomic carbon [CI] and $^{13}$CO(3--2) data were observed with the
APEX telescope\footnote{APEX, the Atacama Pathfinder Experiment is a
  collaboration between the Max-Planck-Institut f\"ur Radioastronomie,
  the Onsala Space Observatory (OSO), and the European Southern
  Observatory (ESO).} in several observing runs between May and July
2018 (project ID M9504A\_101). The approximate area of $0\pdeg25\times 0\pdeg25$ covered in the
on-the-fly mode is outlined by the white box in Figures \ref{8mu},
\ref{hisa} and \ref{large_mom1}. The two-band FLASH receiver was tuned
to the [CI] frequency of 492.160651\,GHz and to the $^{13}$CO(3--2)
frequency of 330.587965\,GHz. Maps were conducted in the on-the-fly
mode doing always maps in Right Ascension and Declination to reduce
potential scanning effects. The data reduction was conducted within
the {\tt GILDAS} framework with the sub-programs {\tt class} and {\tt
  greg} (for more details see {\sf http://www.iram.fr/IRAMFR/GILDAS}).

The original spectral resolution is $\sim$0.05\,km\,s$^{-1}$. However,
to decrease the noise, we binned the data to a spectral resolution of
0.5\,km\,s$^{-1}$. To increase the signal-to-noise ratio for the [CI]
data, we smoothed them to a spatial resolution of $20''$. The
$^{13}$CO(3--2) data are kept at their native spatial resolution of
$18''$. The data are in antenna temperature $T_A^*$. The rms for the
[CI] and $^{13}$CO(3--2) data-cubes in a single 0.5\,km\,s$^{-1}$
channel is 0.14 and 0.12\,K, respectively.

\begin{figure}[htb]
\includegraphics[width=0.49\textwidth]{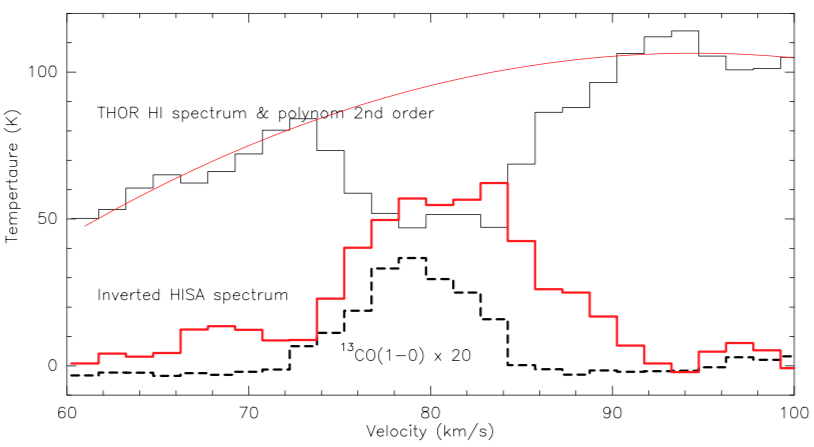}
\caption{Compilation of HI and $^{13}$CO(1--0) spectra for the central area
    of the G28.3 IRDC (averaged over the central $100''$ squared). The
    thin black histogram shows the original HI emission data from the
    THOR survey \citep{beuther2016,wang2020a} where the HI self-absorption
    (HISA) is visible as absorption against the typically bright
    $\sim$100\,K emission of the Galaxy. The red line presents a 2nd
    order polynomial fit to the non-HISA part of the spectrum. The red
    histogram then shows the resulting inverted HISA spectrum for this
    feature. The dashed histogram presents the $^{13}$CO(1--0)
    emission \citep{jackson2006} for the same region (multiplied by 20
    for easier readability).}
\label{spectra} 
\end{figure} 

\section{Results}
\label{results}

\subsection{Large-scale atomic HI and molecular gas $^{13}$CO(1--0) distribution}

The THOR atomic hydrogen data reveal a clear HI self-absorption (HISA)
cloud in the environment of G28.3 (Figs.~\ref{spectra} and
\ref{hisa}). While such a dip in HI emission could also be caused by
missing HI, the fact that this HI emission dip at the given velocities
is correlated with strong $^{13}$CO emission indicates that the lower
HI emission is most likely caused by HI self-absorption (e.g.,
\citealt{riegel1972,heiles1975,vanderwerf1988,gibson2000,kavars2005,denes2018,wang2020b}). Furthermore,
we checked the THOR cm continuum data for background sources against
which we can measure directly the absorption. Although we have only
weak continuum sources in that field prohibiting in-depth real
absorption studies, we do find HI absorption against the continuum at
the respective velocity ranges. This confirms high HI column densities
and hence the HISA interpretation in general G28.3 cloud.

Since HISA features are absorption signatures against the bright HI
emission of the Milky Way, they are tracing the CNM (e.g.,
\citealt{li2003,gibson2000,gibson2005,gibson2005b,wang2020b}). Fitting
a second order polynomial to the channels around the HISA feature, and
inverting the resulting spectra, one can retrieve a HISA spectrum (see
Fig.~\ref{spectra} showing the different parts of that approach). For
more details about the actual HISA extraction process, see
\citet{wang2020b}.

In comparison to the normal HI spectrum with emission at almost all
velocities, such a HISA spectrum has the advantage that it is almost
Gaussian and allows us a much simpler analysis of the CNM than any
normal HI emission spectra would do. Doing this HISA fitting pixel by
pixel, one can derive a ``HISA emission map''. Figure \ref{hisa}
presents this HISA map integrated over the velocity of the cloud
between 71 and 86\,km\,s$^{-1}$. One clearly sees that the general
structure of the CNM traced by the HISA is associated with the
infrared dark cloud G28.3, but that the CNM appears as expected to be
more extended. While the central 875\,$\mu$m emission appears to be
embedded in a larger-scale envelope of CNM, the HISA peak emission is
offset from the 875\,$\mu$m main filamentary structure. This decrease
of HISA towards the main infrared dark filament may be indicative of
ongoing conversion of atomic to molecular gas in the central denser
molecular cloud, also traced by the 875\,$\mu$m emission.

\begin{figure}[htb]
\includegraphics[width=0.49\textwidth]{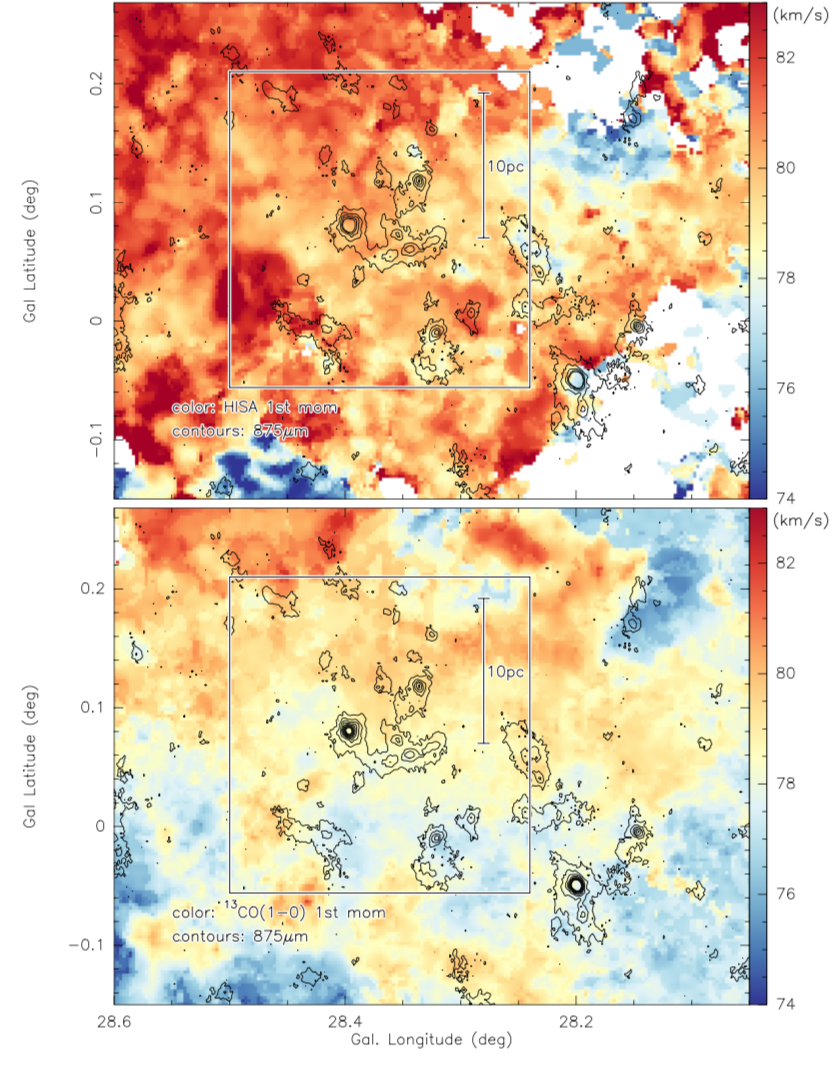}
\caption{Large-scale velocity structure of the cold neutral medium
  (CNM) measured as HISA (top panel) and of the molecular gas measured
  in $^{13}$CO(1--0) (bottom panel, \citealt{jackson2006}). The
  color-scale shows in both panels the 1st moment maps
  (intensity-weighted peak velocities), and the contours show the
  875\,$\mu$m emission from the ATLASGAL survey
  \citep{schuller2009}, contour levels start at $4\sigma$ level of
  200\,mJy\,beam$^{-1}$ and continue in $8\sigma$ steps up to
  1.8\,Jy\,beam$^{-1}$). The white box outlines the approximate region
  imaged with APEX in the [CI] and $^{13}$CO(3--2) emission. Scale
  bars are shown within the box.}
\label{large_mom1} 
\end{figure} 

\begin{figure*}[htb]
\includegraphics[width=0.99\textwidth]{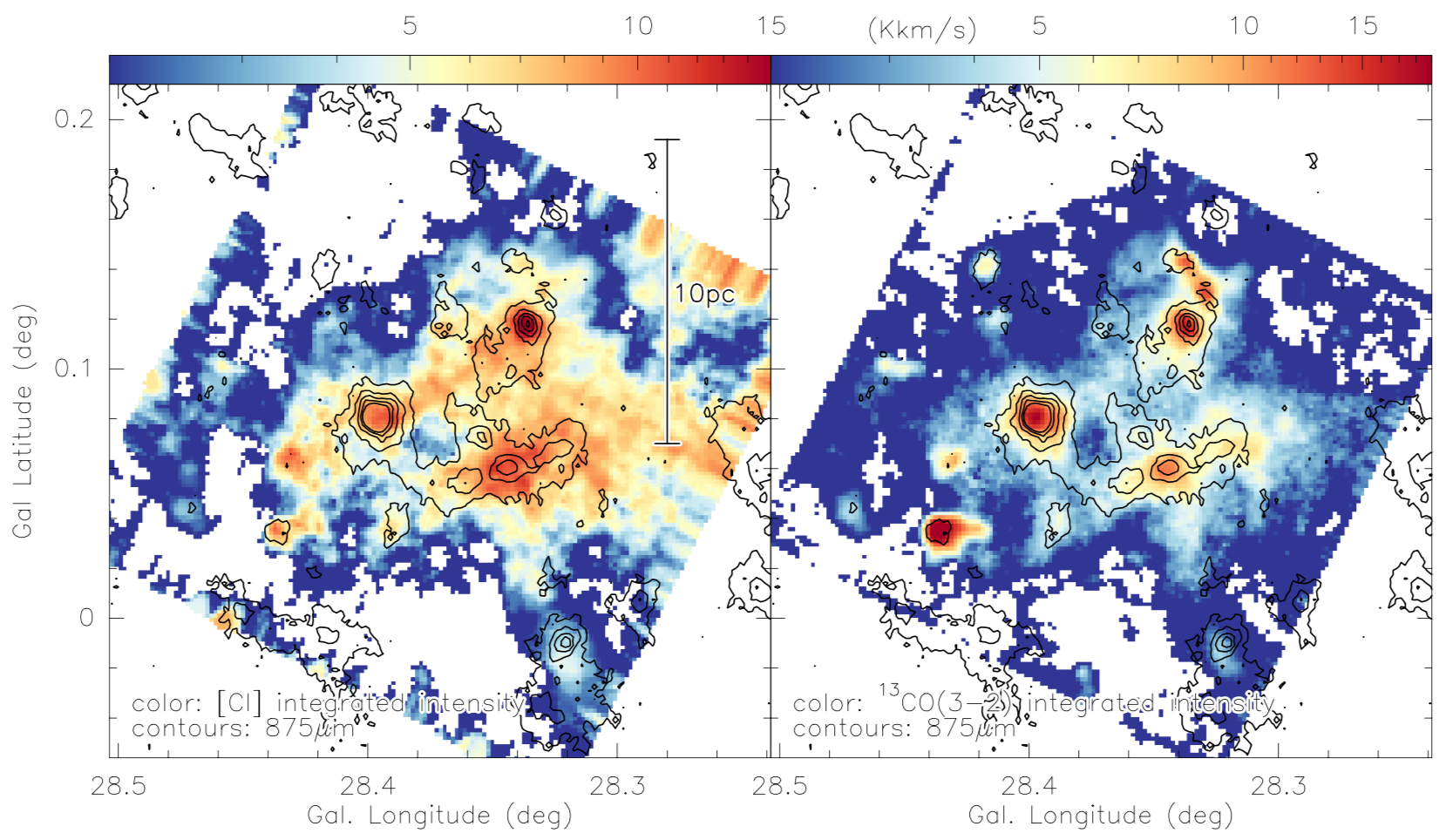}
\caption{Integrated intensity maps of [CI] (left panel) and
  $^{13}$CO(3--2) (right panel). The color-scales present the
  corresponding integrated line maps over a velocity range of
  [71,86]\,km\,s$^{-1}$. The contours show the 875\,$\mu$m emission
  from the ATLASGAL survey (\citealt{schuller2009}, contour levels
  start at $4\sigma$ level of 200\,mJy\,beam$^{-1}$ and continue in
  $8\sigma$ steps up to 1.8\,Jy\,beam$^{-1}$). A scale bar is shown in
  the left panel.}
\label{ci_int} 
\end{figure*} 

Furthermore, one can use the HISA data cube and extract moment maps to
study the velocity structure, similar to typical approaches conducted
with molecular line data. Figure \ref{large_mom1} presents the
corresponding 1st moment maps (intensity-weighted peak velocities) of
the HISA as well as the $^{13}$CO(1--0) data from GRS
\citep{jackson2006}. The general gas velocities of the CNM traced by
the HISA and the molecular gas traced by the $^{13}$CO(1--0) emission
cover the same velocities between approximately 70 and 90\,km\,s
(Figs.~\ref{spectra} \& \ref{large_mom1}). Hence, both should trace
approximately the same large-scale structures in our Milky
Way. However, there are also significant kinematic differences: While
the molecular gas shows a velocity gradient across Galactic latitudes
(similar to the finding of the even denser gas seen in N$_2$H$^+$ on
smaller scales by \citealt{tackenberg2014}) from
$\sim$83\,km\,s$^{-1}$ to $\sim$76\,km\,s$^{-1}$, the HISA 1st moment
map rather shows more uniform velocities around 82--83\,km\,s$^{-1}$
in the outskirts of the cloud with slightly lower velocities toward
the center of the G28.3 region. While this slight shift may be
indicative of a small gradient also in the HISA map, this is
observationally not significant (see also position-velocity discussion
in section \ref{velo}). We will get back to the velocity gradients in
comparison to the [CI] emission in the following section.


%
%
%

\subsection{Velocity structure of atomic and molecular carbon around
  the G28.3 IRDC}
\label{velo}

In the following, we focus on the closer environment around the IRDC
G28.3. Figure \ref{ci_int} presents the integrated [CI] and
$^{13}$CO(3--2) emission from our APEX observations. While both
tracers show strongest emission toward the 875\,$\mu$m peak positions,
the $^{13}$CO(3--2) emission follows the dense gas structures more
closely than the atomic carbon [CI] emission that exhibits a more
diffuse halo-like structure around the central 875\,$\mu$m dust
continuum emission.

\begin{figure*}[htb]
\includegraphics[width=0.99\textwidth]{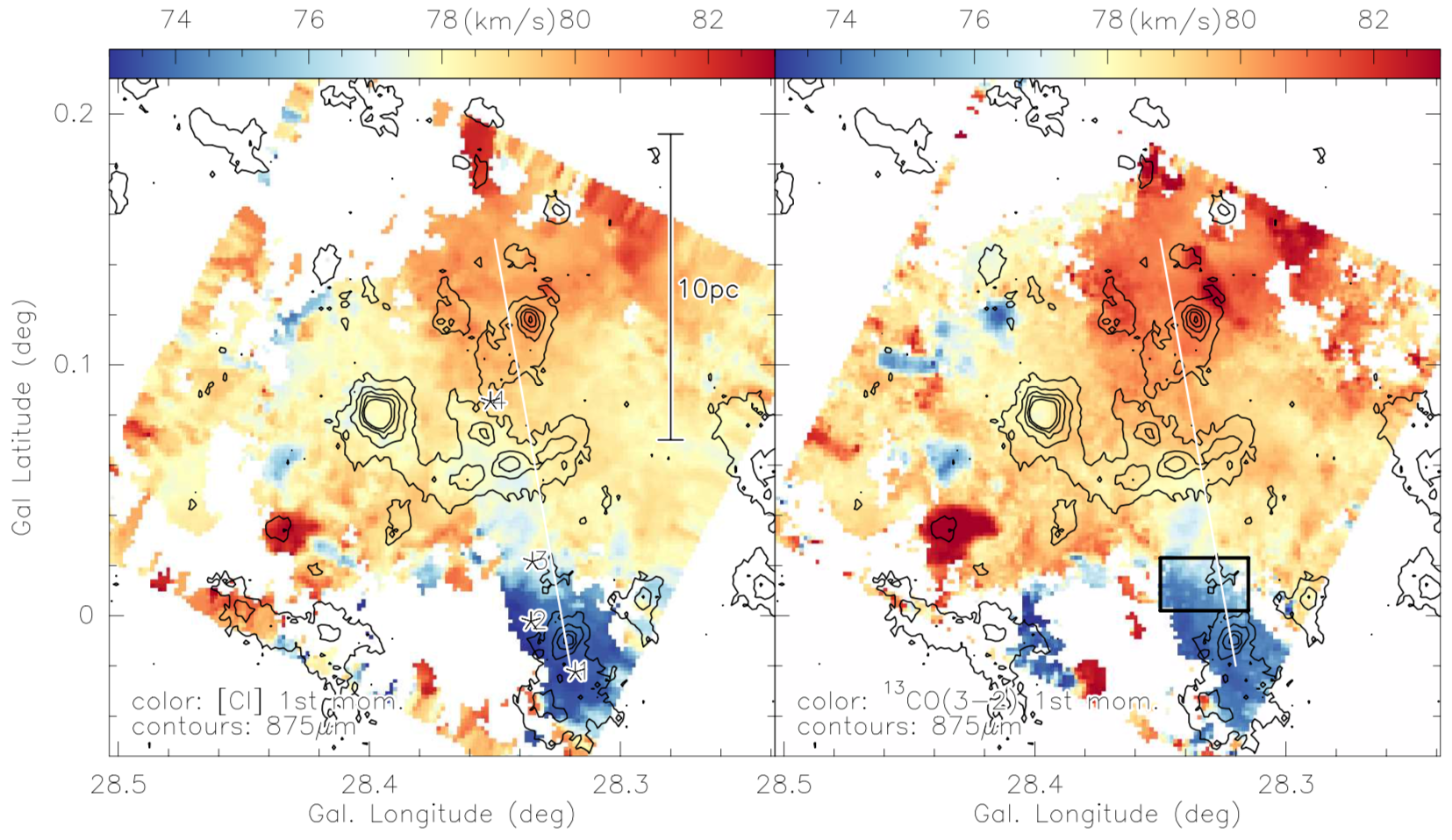}
\caption{First moment maps (intensity-weighted peak velocities) of
  [CI] (left panel) and $^{13}$CO(3--2) (right panel). The
  color-scales present the corresponding integrated line maps over a
  velocity range of [71,86]\,km\,s$^{-1}$. The contours show the
  875\,$\mu$m emission from the ATLASGAL survey
  (\citealt{schuller2009}, contour levels start at $4\sigma$ level of
  200\,mJy\,beam$^{-1}$ and continue in $8\sigma$ steps up to
  1.8\,Jy\,beam$^{-1}$). The white lines show the direction (north to
  south) of the pv-diagrams in Fig.~\ref{pv}. The stars in the left
  panel show the positions of the spectra presented in
  Fig.~\ref{spectra_ci}. A scale-bar is shown in the left panel.
    The little box in the right panel outlines the region with the
    velocity jump.}
\label{moments1}
\end{figure*} 

More interesting than just the integrated emission are the kinematic
signatures found in both tracers. Fig.~\ref{moments1} shows the 1st
moment maps in [CI] and $^{13}$CO(3--2). One can identify in both
tracers a clear velocity gradient from roughly 83\,km\,s$^{-1}$ at
latitudes higher than the central IRDC, and velocities even below
74\,km\,s$^{-1}$ at latitudes below the IRDC. The main dust continuum
filament around a latitude of $\sim 0\pdeg07$ exhibits a rather
uniform peak velocity around $\sim 78$\,km\,s$^{-1}$. whereas the
additional strong 875\,$\mu$m peak at higher latitudes
($\sim 0\pdeg12$) is already more redshifted with velocities
$>80$\,km\,s$^{-1}$. This velocity shift was already identified in the
high-density tracer N$_2$H$^+$ by \citet{tackenberg2014}.

\begin{figure*}[htb]
\includegraphics[width=0.99\textwidth]{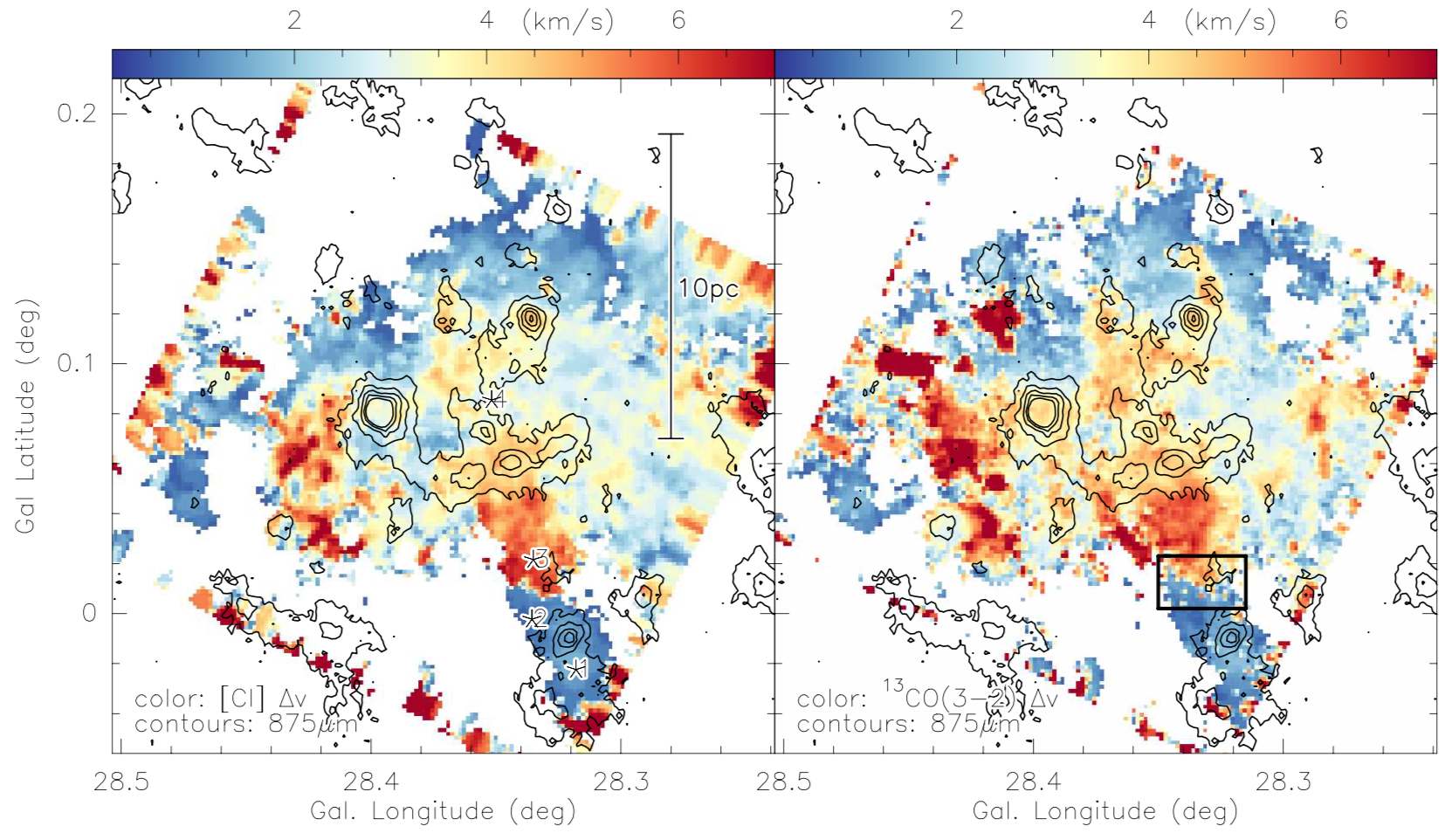}
\caption{Second moment maps (intensity-weighted velocity dispersions)
  of [CI] (left panel) and $^{13}$CO(3--2) (right panel) over a
  velocity range of [71,86]\,km\,s$^{-1}$. The contours show the
  875\,$\mu$m emission from the ATLASGAL survey
  (\citealt{schuller2009}, contour levels start at $4\sigma$ level of
  200\,mJy\,beam$^{-1}$ and continue in $8\sigma$ steps up to
  1.8\,Jy\,beam$^{-1}$). The stars in the left panel show the
  positions of the spectra presented in Fig.~\ref{spectra_ci}. A
  scale-bar is shown in the left panel. The little box in the
    right panel outlines the region with the velocity jump.}
\label{moments2} 
\end{figure*} 

\begin{figure}[htb]
\includegraphics[width=0.49\textwidth]{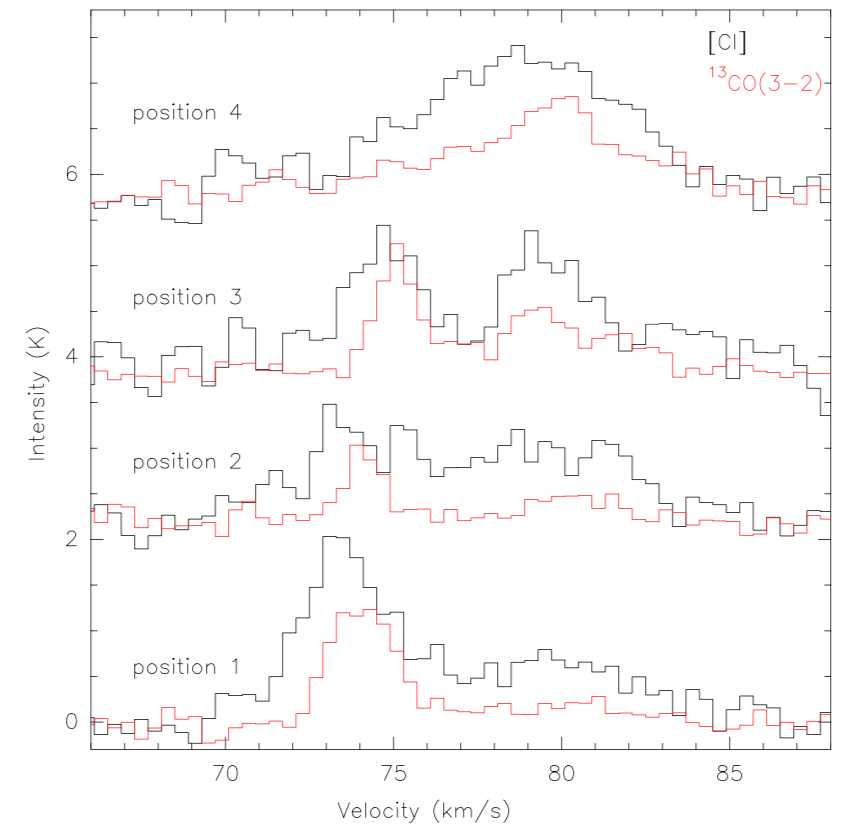}
\caption{[CI] (black) and $^{13}$CO(3--2) (red) spectra toward the
  four positions marked in Figs.~\ref{moments1} and
  \ref{moments2}. The vertical y-axis offsets between the spectra are
  only for better readability.}
\label{spectra_ci} 
\end{figure}

Particularly prominent is a velocity feature at Galactic coordinates
of $\sim 28\pdeg33/0\pdeg01$ where the first moment maps in
Fig.~\ref{moments1} show a strong decrease to even lower velocities,
and at the same position the second moment maps exhibit almost a jump
to low velocity dispersion measurements (see little box in
Figs.~\ref{moments1} and \ref{moments2}). Since moment maps are only
intensity-weighted integral measurements, to investigate this velocity
change in more depth, we extracted individual [CI] and $^{13}$CO(3--2)
spectra toward four positions along that velocity structure marked as
``1'' to ``4'' in Figures \ref{moments1} and \ref{moments2}. These
spectra are shown in Figure \ref{spectra_ci}.

The [CI] and $^{13}$CO(3--2) spectra toward that region clearly
exhibit several velocity components. While the southernmost position
``1'' shows mainly one velocity component around 73-74\,km\,s$^{-1}$,
the northernmost position ``4'' is dominated by a (broader) component
at around 79\,km\,s$^{-1}$. The two positions inbetween, and
particularly prominent toward position ``3'' show two component spectra
combining the features seen individually toward positions ``1'' and
``4''.

Having two different velocity components in the environment of that
IRDC raises the question whether we are witnessing the potential
interaction of two gas flows. These could be either externally driven
colliding flows and/or cloud-cloud collisions, or they could be driven
by self-gravity, that may trigger the formation of a dense
star-forming molecular cloud at its converging point (e.g.,
\citealt{duarte2011,bisbas2017,inoue2018,kobayashi2018,haworth2015,haworth2018,vazquez2006,heitsch2008,banerjee2009,gomez2014,vazquez2019}). Regarding
the externally driven processes, one can consider the cloud-cloud
collisions as a sub-group of the more general colliding gas flows. The
main difference should be the state of the gas: While colliding flows
are in general considered to be continuous gas flows that start as
more diffuse HI clouds with a mix of CNM and WNM, the cloud-cloud
collision picture typically refers to more concrete objects consisting
mainly of cold, molecular gas (e.g.,
\citealt{haworth2015,haworth2018,bisbas2017}).

\begin{figure*}[htb]
\includegraphics[width=0.99\textwidth]{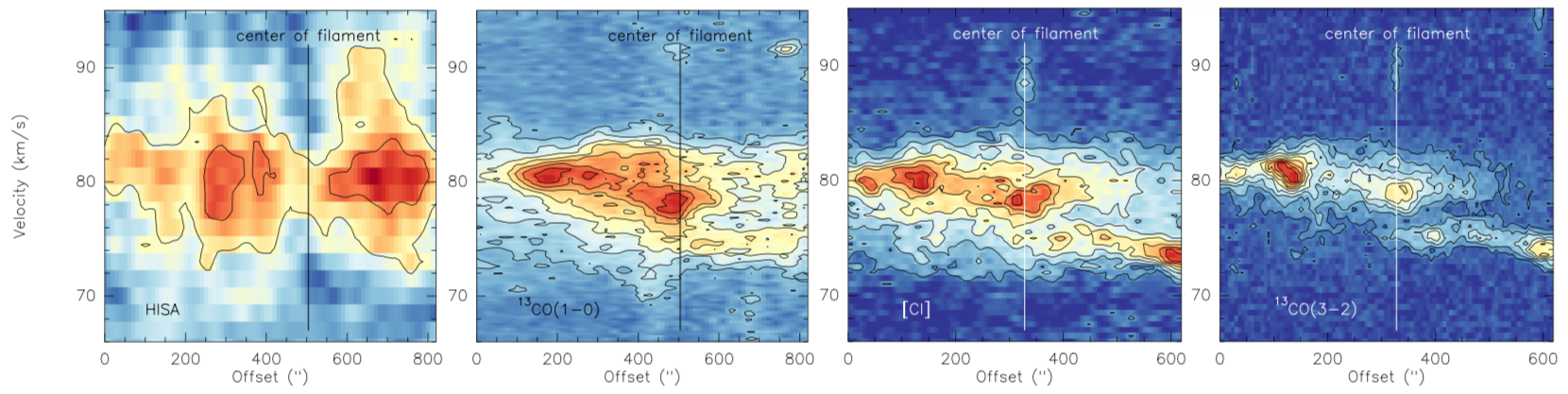}
\caption{Position-velocity diagrams for the four tracers from left to
  right: HISA, $^{13}$CO(1--0), [CI] and $^{13}$CO(3--2). The cuts are
  in a north-south direction along the lines shown in
  Figs.~\ref{moments1} ([CI], $^{13}$CO(3--2)) and \ref{13co10_zoom}
  (HISA, $^{13}$CO(1--0)). Both axes have slightly larger extent for
  the HISA and $^{13}$CO(1--0) data compared to the [CI] and
  $^{13}$CO(3--2) cuts. The approximate location of the center of the
  filament is marked in all panels.}
\label{pv} 
\end{figure*}

\begin{figure}[htb]
\includegraphics[width=0.49\textwidth]{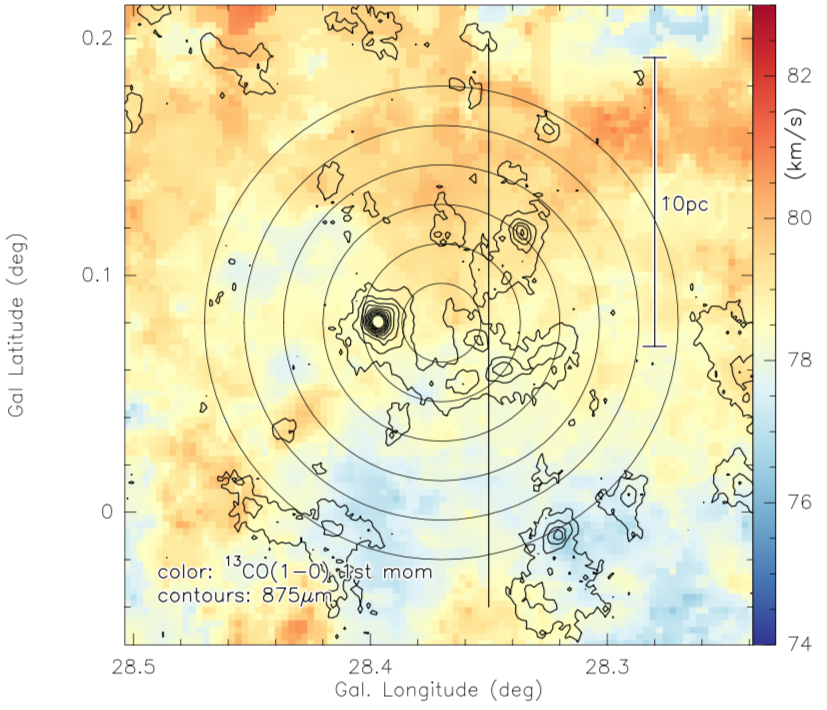}
\caption{Zoom into the $^{13}$CO(1--0) 1st moment map from
  Fig.~\ref{large_mom1}. The color-scale shows the 1st moment map, and
  the contours present the ATLASGAL 875\,$\mu$m emission
  \citep{schuller2009}, contour levels start at $4\sigma$ level of
  200\,mJy\,beam$^{-1}$ and continue in $8\sigma$ steps up to
  1.8\,Jy\,beam$^{-1}$). A scale bar is shown. The black line presents
  the direction (north to south) of the pv-diagrams in
  Fig.~\ref{pv}. The circles outline annuli with radii starting at
  $60''$ and then increasing in $60''$ steps to $360''$ (for details
  see section \ref{flow}).}
\label{13co10_zoom} 
\end{figure}

To better evaluate such potential converging gas flows, Figure
\ref{pv} presents position-velocity cuts along approximate north south
direction in all four gas tracers, HI, $^{13}$CO(1--0), [CI] and
$^{13}$CO(3--2). The corresponding cut directions for the two pairs of
tracers (HI/$^{13}$CO(1--0) for the larger scales, and
[CI]/$^{13}$CO(3--2) for the closer environment of the IRDC) are shown
if Figs.~\ref{13co10_zoom} and \ref{moments1}, respectively. While for
the larger-scale HISA and $^{13}$CO(1--0) data, we select the most
straightforward north-south direction, for the [CI] and
$^{13}$CO(3--2) emission, the cut is slightly inclined ($\sim$10$^0$,
Fig.~\ref{moments1}) to follow the larger extent of the [CI] and
$^{13}$CO(3--2) emission along that axis. We checked more orientations
on the different data, and this small difference does not change the
results discussed below. The orientation of the pv-diagrams is
north-south, i.e., offset $0''$ is the northern end of the cut.

These pv-diagrams exhibit a few interesting features. To start with
the diffuse emission, the HISA pv-diagram shows no obvious velocity
gradient over the whole extent of $\sim 800''$ (roughly 18.25\,pc at
4.7\,kpc distance). Going to the molecular gas, the $^{13}$CO(1--0)
cut peaks in the north at roughly 80.5\,km\,s$^{-1}$, stays in that
regime for about $300''$, and then exhibits a gradient toward the
center of the filament to $\sim$77.5\,km\,s$^{-1}$. Of particular
interest is then also the regime south of the main filament where the
$^{13}$CO(1--0) emission shows two components, one again around
80.5\,km\,s$^{-1}$, and the second component at lower velocities of
$\sim$74.5\,km\,s$^{-1}$.

The two pv-diagrams for [CI] and $^{13}$CO(3--2) cover a slightly
smaller length with about $600''$ or $\sim$13.7\,pc. Around the
central filament, they exhibit similar signatures with velocities
around 80.5\,km\,s$^{-1}$ in the north and even below 74\,km\,s$^{-1}$
in the south. The filament itself shows again intermediate velocities
between 77 and 79\,km\,s$^{-1}$. While the overall signatures for [CI]
and $^{13}$CO(3-2) are similar, [CI] shows a bit more extended
emission with a larger velocity spread (see also
Fig.~\ref{ci_int}). The division into separate velocity components is
most prominent in the highest-density tracer $^{13}$CO(3-2). With the
higher spatial and spectral resolution of the [CI]/$^{13}$CO(3--2)
data compared to the $^{13}$CO(1--0) data, one also more clearly resolves
the gradient-like velocity structure of the two components from the
north and the south toward the central filament.

\begin{figure}[htb]
  \includegraphics[width=0.49\textwidth]{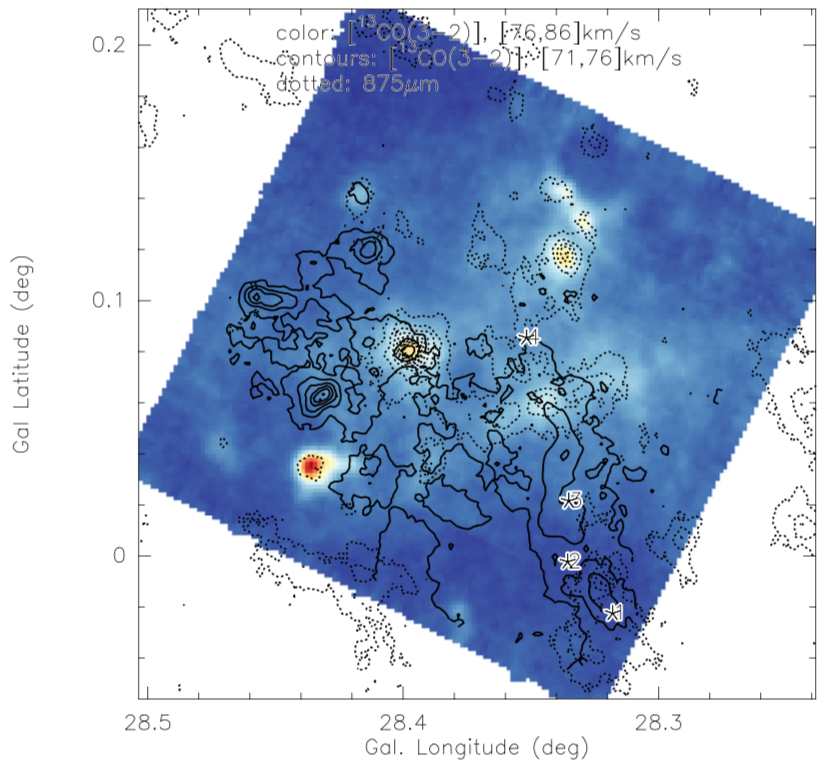}
\caption{Integrated $^{13}$CO(3--2) emission for two velocity
  components. The color scale and full contours show the [76,86] and
  [71,76]km\,s$^{-1}$ components, respectively. The contour levels are
  in $9\sigma$ steps (with $1\sigma=0.125$\,K\,km\,s$^{-1}$). The
  dotted contours show the corresponding 875\,$\mu$m continuum
  emission from \citet{schuller2009} as reference frame. The stars
  again show the positions of the spectra presented in
  Fig.~\ref{spectra_ci}.}
\label{2components} 
\end{figure} 

For visualization purpose, Fig.~\ref{2components} shows the
$^{13}$CO(3--2) integrated emission maps for the two velocity regimes
[71,76]\,km\,s$^{-1}$ and [76,86]\,km\,s$^{-1}$, respectively. Although
  there is no clear north-south separation of the two components, the
  higher-velocity gas is preferentially found toward the north and the
  lower-velocity gas more toward the south. One has to keep in mind
  that these gas components and kinematic signatures are always
  projections onto the plane of the sky, and therefore, clear spatial
  separations would even be a surprise.

We note that while the first moment maps of $^{13}$CO(1--0) and
[CI]/$^{13}$CO(3--2) (Figs.~\ref{large_mom1} and \ref{moments1}) give
the visual impression that the velocity gradient may increase with
critical density of the tracer, this is most likely mainly caused by
the two velocity components and the fact that the second component
around 74.5\,km\,s$^{-1}$ is more pronounced in the higher density
tracers [CI]/$^{13}$CO(3--2). Closer inspection of the pv-diagrams in
Fig.~\ref{pv} for these three tracers, considering only the transition
from the north to the center of the filament, show that there is no
large difference in the magnitude of the velocity gradient. Similar
velocity regimes were also found in the even higher-density tracer
N$_2$H$^+$(1--0) by \citet{tackenberg2014}.

Combining these different velocity signatures, the data show strong
signatures of two gas components at different velocities (around
6\,km\,s$^{-1}$ apart) that converge to a common intermediate velocity
at the location of the infrared dark cloud and active star-forming
region, similar to filament formation via gravitationally driven,
converging gas flows (e.g., \citealt{gomez2014}). We interprete these
signatures as indicators of converging gas streams that may trigger
the star formation event at its center (e.g.,
\citealt{vazquez2006,heitsch2008,banerjee2009,gomez2014}).
Position-velocity diagrams based on simulations of cloud-cloud
collisions sometimes show a characteristic pattern of lower-level
emission between two main velocity components, a so-called bridging
feature (e.g., \citealt{haworth2015,haworth2018}). Similar signatures
were also reported in observations (e.g.,
\citealt{jimenez-serra2010,henshaw2013,dobashi2019,fujita2019}). The
pv-diagrams of the G28.3 region presented here (Fig.~\ref{pv}) show
different signatures in the sense that there is not a lower-intensity
bridge between the two well-defined components, but that the two
velocity components converge at the center of the cloud toward a
central, high-intensity velocity component. However, the absences of a
bridging feature does not necessarily rule out the formation of G28.3
in a cloud-cloud collision, as this feature is not always visible in
simulations. For example, simulations by \citet{bisbas2017} show that
the low-intensity bridge feature may merge into a centrally peaked
pv-diagram during the evolution of the cloud-cloud collisions while
the colliding-cloud simulations of \citet{clark2019} yield only a
single central velocity component in CO or [CI], with multiple
components only becoming apparent when one looks at the [CII] emission
from the cloud. In addition to this, the multiple components and
bridging feature may not be visible in cases where our line of sight
is oriented at a large angle to the direction of motion of the
clouds. We get back to the interpretation in sections
\ref{simulations} and \ref{dynamicalSF}.

\section{Discussion}
\label{discussion}

\subsection{Histogram of Oriented Gradients (HOG)}
\label{sectionHOG}

A way to evaluate similarities in the structures traced by the
different spectral lines is the histogram of oriented gradients (HOG),
a statistical method from machine vision recently introduced in
astrophysical data analysis by \citet{soler2019}. Basically, the HOG
analysis compares intensity maps by comparing the relative orientation
between its gradients.  The degree of correlation between the images
is estimated using the projected Rayleigh statistic ($V$), which is a
test that quantifies if the relative angle distribution is flat, as it
is the case of two completely uncorrelated maps, or peaked around
0\,deg, as it is the case of two maps with a significant degree of
correlation.

\begin{figure}[htb]
\includegraphics[width=0.49\textwidth]{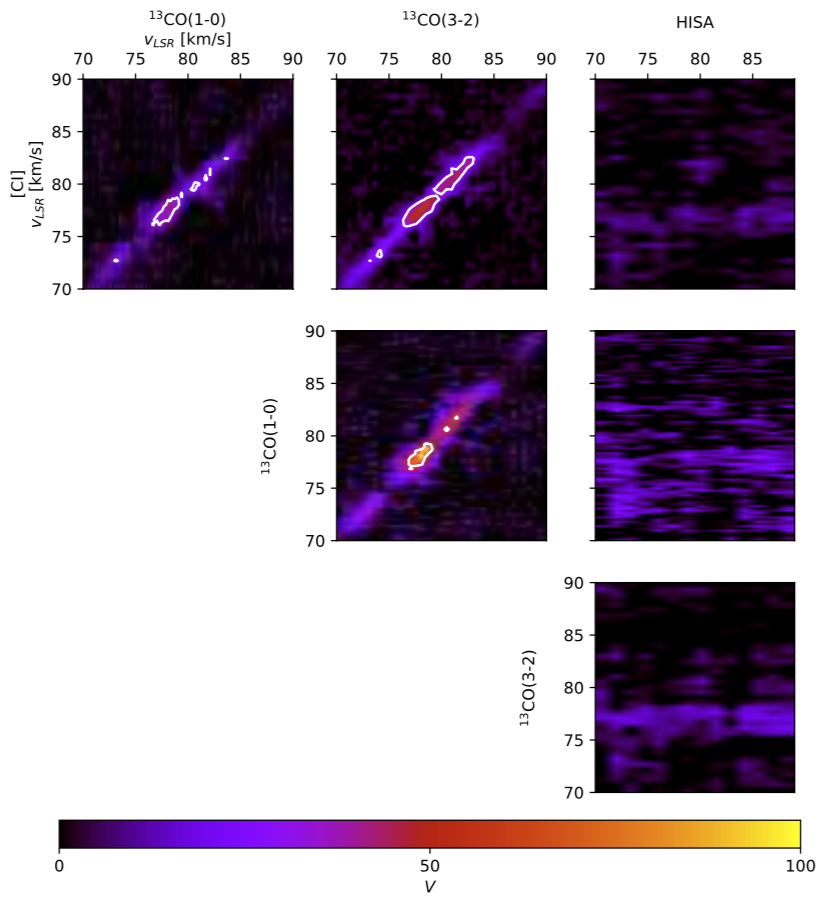}
\caption{Spatial correlation between the four different tracers across
  velocity channels as evaluated using the histogram of oriented
  gradients method (HOG, \citealt{soler2019}).  The figure shows all
  tracers correlated with each other as labeled at each plot. The
  panels show the values of the projected Rayleigh statistic ($V$), a
  measure of the morphological correlation between velocity channels.
  Large values of $V$ indicate large correlation between the
  corresponding velocity-channels maps.  The contours correspond to
  the 7\,$\sigma$ confidence limit on the $V$ values, see
  \citet{soler2019} for more details.}
\label{hog} 
\end{figure}

The HOG can be used to compare individual single-frequency maps, but
also to position-position-velocity cubes.  In that case, the result of
the analysis is a matrix of $V$ values for the different velocity
channels in each tracer, which is shown in Fig.~\ref{hog}. We use this
matrix of $V$ values to investigate whether the gas traced by
different spectral lines follows a similar spatial distribution across
velocity channels. \citet{soler2019} applied the HOG analysis to
spectral line data as well as different simulations, and they found,
for example, that constant converging gas flows result in spatial and
kinematic structures agreeing well between different spectral lines
(mainly showing high projected Rayleigh-V values along the diagonal in
plots like those in Fig.~\ref{hog}), whereas feedback processes can
produce kinematic offsets between, e.g., atomic and molecular gas
(showing high projected Rayleigh-V values offset from the diagonal). For
more details about the method and its implications, please see
\citet{soler2019}.

We applied this HOG method to the four datasets studied here,
namely, [CI], $^{13}$CO(1-0), $^{13}$CO(3-2), and HISA
position-position-velocity cubes.  The resulting correlations,
quantified by $V$, are presented in Fig.~\ref{hog}.  We find that the
HISA does not show significant correlation with any of the other
tracers across the velocity range between 70 and 90\,km\,s$^{-1}$.  In
contrast to this, the other three tracers show high spatial
correlation across velocity channels, evidenced by high $V$ values.
The largest values of $V$ are concentrated along the diagonal of the
correlation plot, that is, across the same line-of-sight velocity in
each pair of tracers.  This result is expected for tracers that are
co-spatial and co-moving.

The low values of $V$ found in the comparison between the HISA and the
other tracers indicate that there is no morphological correlation
between CNM as traced by HISA and the other denser gas tracers across
all measured velocity channels.  This lack of correlation can be
interpreted as an indication of the decoupling of the dense gas
of the main cloud from the mostly atomic CNM in the more diffuse cloud
envelope. This is slightly different to the morphological correlation
found between HI and $^{13}$CO(1--0) emission in the molecular clouds
studied in \citet{soler2019}. It appears that correlation or
non-correlation between HI and denser gas tracers may not be universal
but could be an indication of different evolutionary stages in this
kind of objects. We note that for the G28.3 cloud studied here, the
HISA itself is also weaker toward the central filamentary IRDC as
visible in Figs.~\ref{hisa} and \ref{pv}. If one interpretes this
depression of HISA towards the center as a sign of potential gas
conversion from atomic H to molecular H$_2$, this can be taken as
further support for the kinematic decoupling of the CNM, as traced by
HISA, and the molecular phase.

\subsection{Velocity structure functions}

A different way to investigate the kinematics of molecular clouds is
the analysis of velocity structure functions (e.g.,
\citealt{miesch1994,ossenkopf2002,esquivel2005,heyer2015,chira2019};
Henshaw et al.~subm.). Velocity structure functions are essentially a
measure of how much the velocity measured between pixels separated by
a given distance varies as a function of spatial scale. Mathematically
the structure function $S_p$ of order $p$ is described as:

$$ S_p(l) = <|v(r)-v(r+l)|^p> $$

where $l$ is known as the spatial lag and represents the separation
between pairs of points $S_p$ is calculated for.  As outlined for
example in \citet{chira2019}, the slope of the structure function can
be used to infer whether turbulence and/or gravity play an important
role in shaping the gas dynamics of the cloud.  \citet{chira2019} show
in their simulations that purely turbulence-dominated structure
functions are steeper than those where gravity becomes more important
because gravity increases the velocity on small scales so that the
structure function becomes shallower.

We now derived the velocity structure function $S_2$ of order 2 for
all four gas tracers from their corresponding 1st moment maps
(intensity-weighted peak velocities). To allow a meaningful
comparison, the four 1st moment maps were smoothed to the $46''$
spatial resolution of the $^{13}$CO(1--0) data and all put on the same
pixel grid. No large-scale gradient was removed from the data. Since
we derived the structure function from the 1st moment maps, and these
are intensity-weighted peak velocities, the derived structure
functions can be considered as (column-)density weighted velocity
structure functions.

\begin{figure}[htb]
\includegraphics[width=0.49\textwidth]{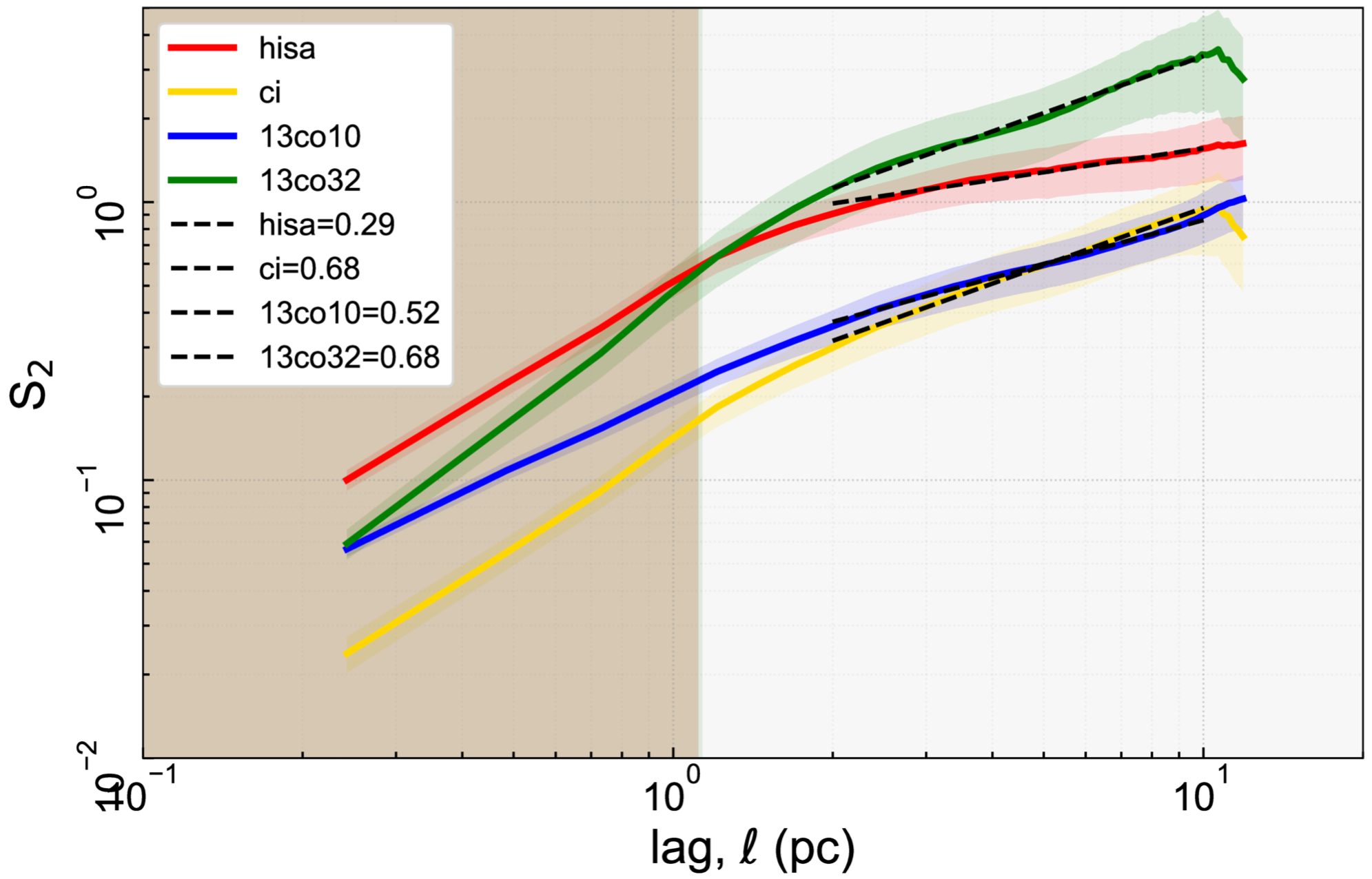}
\caption{One-dimension velocity structure function. The structure
  function $S_2$ is plotted versus the spatial lag $l$ for all four
  observed gas tracers. The red, blue, yellow and green lines
  correspond to the HISA, $^{13}$CO(1--0), [CI] and $^{13}$CO(3--2)
  data, respectively. The grey-shaded part is the spatial-resolution
  limit corresponding to the $46''$ beam of the $^{13}$CO(1--0)
  data. The slopes are fitted outside that limit and shown as dashed
  lines. The corresponding fit results are presented in the legend.}
\label{s1d} 
\end{figure} 

Figure \ref{s1d} presents the velocity structure function $S_2$
for all four tracers plotted against the spatial lag $l$.  While the
$^{13}$CO(1--0) and [CI] structure functions resemble each other well
not just in the slope but also in the magnitude of the velocity
fluctuations $S_2$, the HISA and $^{13}$CO(3--2) structure functions
exhibit slightly elevated values of $S_2$. For $^{13}$CO(3--2), this
is more intuitively clear because one sees very different velocities
in the north and south of the region (e.g., Figs.~\ref{moments1} \&
\ref{pv}). The high absolute values of $S_2$ for the HISA data,
combined with the flat slope, imply that there are comparably large
velocity differences, but that these do not change strongly with
spatial scale as indicated by the shallower slope.

Important information about the kinematics of the gas are encoded in
the slopes of the velocity structure functions. The overall regimes of
derived slopes between 0.29 and 0.68 is well within the regime of
other observational studies (see, e.g., the compilation in Table 1 of
\citealt{chira2019}). However, while the slopes are similar for
$^{13}$CO(1--0), [CI] and $^{13}$CO(3--2) (values between 0.52 and
0.68), the slope of the HISA velocity structure function is flatter
with a value of 0.29.

One should keep in mind that other effects like noise or varying
optical depth can also affect the slope of the structure function
(e.g., \citealt{dickman1985,bertram2015}). While optical depth should
not be a big issue for the used tracers (e.g.,
\citealt{shimajiri2013,riener2020}), the flatter slope of the HISA
structure function needs a bit more attention considering potential
noise effects. In principle, noise can introduce a flattening of a
structure function, however, this is most severe on small spatial lags
(e.g., recent simulations in Henshaw et al.~sub.), and we only start
fitting the structure function at spatial lags clearly above the
spatial resolution (Fig.~\ref{s1d}), limiting this effect
already. Furthermore, the 1st moment maps are derived while clipping
values below an rms thresholds to avoid fitting the noise. Taking the
HISA data, the original rms of $\sim$10\,mJy\,beam$^{-1}$ corresponds
to a brightness temperature rms of $\sim$4\,K. The velocity structure
function shown in Fig.~\ref{s1d} is derived with clipping all data
below 15\,K. To check whether the clipping affects the results, we
tested the analysis with a much more conservative clipping threshold
of twice the previous value, i.e., 30\,K. The outcome of that test is
that within the fitting errors, the slope of the HISA velocity
structure function stays the same. Therefore, we infer that noise is
unlikely to explain the flatter slope of the HISA velocity structure
function compared to the other tracers.

While the absolute values of the slopes may be affected by some of the
issues discussed above, the relative difference of almost a factor 2
in slope between the HISA and the other tracers appears to be a solid
result. In the context of the HOG analysis in the previous chapter
(section \ref{sectionHOG}), the slope difference between HISA and the
other tracers may be considered as further support for the decoupling
of the CNM, as traced by HISA, from the denser molecular gas. The
flatter HISA structure function slope indicates that the magnitude of
the velocity fluctuation $S_2$ changes less with varying length scale
than it is the case for the other tracers.

Comparing the slopes of observational velocity structure functions
(compiled from literature) with their simulations, \citet{chira2019}
conclude that the observed clouds are consistent with an intermediate
evolutionary stage that are neither purely turbulence dominated (e.g.,
driven by supernovae remnants) nor gravitationally collapsing, but
where the gas flows are dominated by the formation of hierarchical
structures and cores. The cloud we are observing here around the
infrared dark cloud G28.3 appears to be in a similar evolutionary
stage.
  
\subsection{Mass flows}
\label{flow}

With the given velocity structure, we like to derive estimates for the
mass flow rates of the gas. In a simple form, one can approximate the
mass flow rate $\dot{M}$ in units of $\rm{M}_{\odot}\rm{yr}^{-1}$ via

$$\dot{M} = \Sigma \cdot \Delta v \cdot r  \,\,\, [\rm{M}_{\odot}\rm{yr}^{-1}]$$

with the column density $\Sigma$, a velocity difference $\Delta v$ and
the radius $r$. Since $^{13}$CO(1--0) emission is known to be either
optically thin or to exhibit only comparably low optical depth (see,
e.g., a detailed analysis in \citealt{riener2020}), we chose the
$^{13}$CO(1--0) map to estimate $\Sigma$. Furthermore, $^{13}$CO(1--0)
peak brightness temperatures are at most 2-3\,K in the G28.3
cloud. Comparing these to typical IRDC temperatures between 15 and
20\,K (e.g., \citealt{wienen2012,chira2013}), this is additional
support for the optically thin assumption. The H$_2$ column densities
$\Sigma$ were then estimated from mean $^{13}$CO(1--0) integrated
intensities within the annuli. We applied standard column density
calculations (e.g., \citet{rohlfs2006}) using a $^{13}$CO-to-H$_2$
conversion factor following \citet{frerking1982} and a uniform
excitation temperature of 15\,K. To get a global view of the mass
flow, we use circular annuli with radii starting at $60''$ from the
center and then increase in $60''$ steps out to $360''$ (see
Fig.~\ref{13co10_zoom}). The circularly averaged H$_2$ column
densities range between 4.3 and $6.3\times 10^{21}$\,cm$^{-2}$.

For the velocity difference $\Delta v$ we use the mean difference of
the peak velocities measured in the $^{13}$CO(1--0) 1st moment map at
the edges of the annuli in the north and south, respectively
(Fig.~\ref{13co10_zoom}). This $\Delta v$ measurement is obviously
affected by line-of-sight projection and depends on the inclination of
any potential gas flow. It should hence be considered as a lower
limit. As radius we use the extent of the annuli of $60''$. While each
of the selections can be debated, the results are just meant to give a
rough estimate of the flow rates over the extent of the cloud.

\begin{figure}[htb]
\includegraphics[width=0.49\textwidth]{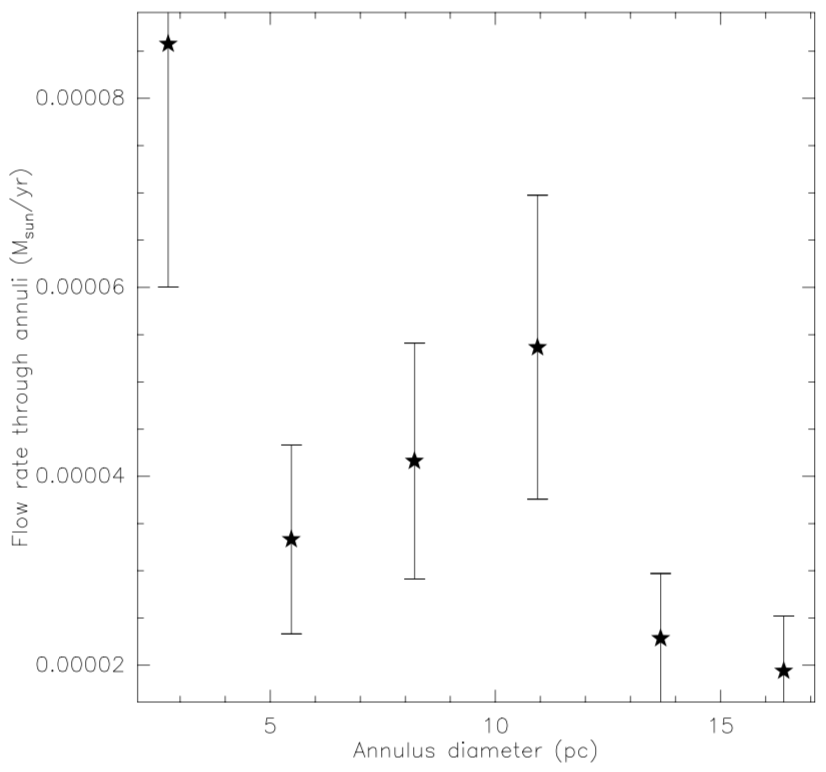}
\caption{Estimated mass flow rate versus the distance of the chosen
  annulus from the cloud center (see Fig.~\ref{large_mom1}). Error-bars
  are just approximate $\pm$30\% ranges.}
\label{flowrate} 
\end{figure} 

Employing this approach, we can estimate the mass flow rate at varying
distances from the central filament. Figure \ref{flowrate} presents
the derived $\dot{M}$ values versus the distance of the annulus from
the center of the region. Being on the order of a few times
$10^{-5}$\,M$_{\odot}$yr$^{-1}$, these flow rates appear rather
constant over the extent of the cloud. The absolute values should be
considered as lower limits because of the two-dimensional projection
of the velocity gradient on the plane of the sky. Interestingly, in
the gravitationally driven cloud collapse simulation by
\citet{gomez2014}, an accretion rate onto the central filament of
$\sim$150\,M$_{\odot}$(Myr)$^{-1}$, corresponding to
$1.5\times 10^{-4}$\,M$_{\odot}$yr$^{-1}$, is inferred. Taken into
account that our observed values should be lower limits because of
projection effects, both accretion rates appear to agree within an
order of magnitude. However, one should keep in mind that this
comparison is based on only one cloud and one simulation, so more
statistical work is needed on both sides to infer tighter constraints
on the mass flow rates during cloud formation in general.

These data can be interpreted as indicative for a constant mass flow
from large to small spatial scales. Such constant flow rates could,
for example, occur in a self-similar gravitationally dominated
collapse picture (e.g., \citealt{whitworth1985,li2018b}).

%
%
%
%

\begin{figure}[htb]
\includegraphics[width=0.4\textwidth]{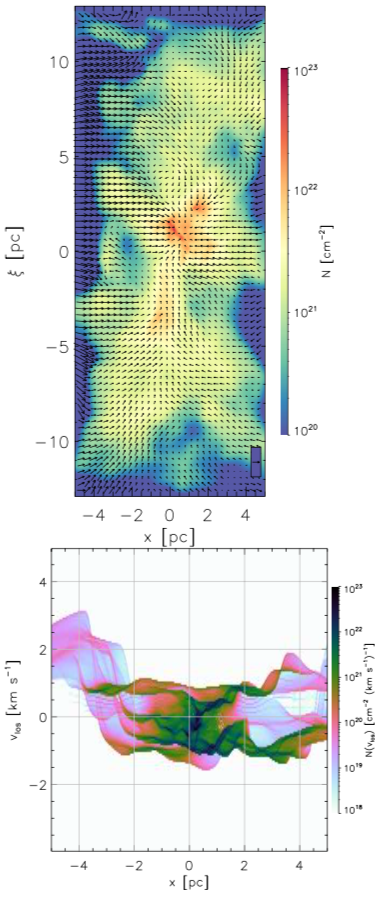}
\caption{Simulation results corresponding to data first published in
  \citet{gomez2014}. The top panel presents a column density
  projection of the simulations at a time step of 21.5\,Myr with
  velocity vectors showing the gas motions on top. The bottom panel
  shows the corresponding position-velocity cut along the X-axis at a
  Y-offset of 0.1\,pc. The color-coding in both plots shows the column
  density of the gas.}
\label{sim} 
\end{figure}

\subsection{Comparison with simulations}
\label{simulations}

As one possible comparison of our observational results with numerical
modeling studies, we choose the molecular cloud formation simulation
originally presented in \citet{gomez2014}. They model the formation of
the cloud of comparable size to G28.3 via the collision of two
oppositely directed warm gas streams (Fig.~\ref{sim}). The collision
triggers a phase transition to the cold medium, the cloud grows
rapidly in mass until it becomes Jeans-unstable and then
collapses. The successive contracting motions of the cloud are
gravitationally driven. The initial low-density cloud setup of the
simulation encompasses an area with diameter of $\sim$50\,pc, and they
form dense filaments of $\sim$15\,pc length with masses of
$\sim$600\,M$_{\odot}$ above densities of $10^3$\,cm$^{-3}$. These
values agree order-of-magnitude-wise with our observed cloud and
filamentary region G28.3. The collapse is simulated in the
smoothed-particle hydrodynamics framework with the GADGET-2 code
including self-gravity and heating and cooling functions that imply
thermal bistability of the gas. For more details about the
simulations, we refer to \citet{gomez2014}.

We are focusing here on their ``Filament 2''. While the paper mainly
shows the images from a snapshot at 26.6\,Myr of their modeled
evolution, here we present a slightly younger evolutionary stage at
21.5\,Myr. At this time, the filamentary cloud is still wider and the
central clump less filamentary, although it is already undergoing
cloud-scale gravitational contraction, accreting from the background
mostly in the direction perpendicular to the filament, and along the
filament onto the central hub. This can be seen in Fig.~\ref{sim} (top
panel), which shows a column density projection of the cloud and
velocity vectors on the plane of the figure.

Even more important are the kinematics, and Figure \ref{sim} (bottom
panel) shows a position-velocity perpendicular to the filamentary
structure. In analogy to our observations, the simulation X-axis
resembles the north-south orientation of our IRDC G28.3
data. Interestingly, at negative offsets, the gas peaks at velocities
of roughly 2.5\,km\,s$^{-1}$ whereas at positive offsets most of the
gas is at around -0.5\,km\,s$^{-1}$. It should be noted that at these
positive offsets there remains still some lower column density gas
left at positive velocities. While the magnitude of the velocity
difference between the two sides of the cloud is smaller than in our
observations, qualitatively, the position-velocity cuts of the
simulations and observations resemble each other well.

Since here we are comparing only one simulation with one observational
dataset, deriving general conclusions is difficult. However, the
qualitative agreement between the observations and simulations
indicates that the G28.3 cloud may indeed form via gravitationally
driven converging gas flows.

Another, maybe more speculative aspect of these simulations is that at
large scales, a phenomenon similar to the inside-out collapse operates
(see Sec.~7.2 of \citealt{vazquez2019}). That is, there seems to be an
expanding infall wave, so that material further outside begins to fall
in later. We speculate that this behavior could also explain the
apparent decoupling of the HI from the denser molecular gas in our
observational data.

\subsection{Dynamical star formation}
\label{dynamicalSF}

How can we interprete the various results obtained above in the
framework of dynamical star formation? Synthesizing the results from
the previous sections, we find:

\begin{itemize}

\item The dense infrared dark cloud is embedded in an extended cloud
  of more diffuse molecular and atomic gas. While the CNM traced by
  the HISA does not show any clear velocity gradient, the other three
  tracers $^{13}$CO(1--0), [CI] and $^{13}$CO(3--2) exhibit a velocity
  gradient from north to south. The HOG and velocity structure
  function analysis confirm this difference indicating a decoupling of
  the atomic CNM from the denser central cloud. This is further
  supported by the decrease of HISA towards the central filamentary
  part of the cloud.

\item The position-velocity diagrams of $^{13}$CO(1--0), [CI] and
  $^{13}$CO(3--2) reveal two velocity components from the north and
  the south that converge at a velocity of $\sim$78\,km\,s$^{-1}$ of the
  central infrared dark cloud. This is indicative of a converging gas
  flow.

\item Estimates of the gas flow rate from the cloud edge to the center
  reveal rather constant values over all scales. This can be
  interpreted as a constant gas flow from the outside to the central
  infrared dark cloud. Such constant flow rate is consistent with a
  self-similar, gravitationally driven collapse of a cloud.

\item Comparing the derived second order velocity structure
  functions with those derived from simulated molecular clouds, the
  data are consistent with gas flows that are dominated by the
  formation of hierarchical structures and cores.

\end{itemize}

Taking these results together, the spatial and kinematic signatures
obtained toward the infrared dark cloud G28.3 are consistent with
converging gas flows that may trigger the formation of the central
IRDC and the gravitational collapse of the cloud as site of active
star formation. It is difficult to clearly differentiate between
colliding flows that started in the low-density medium, or a
cloud-cloud collision picture where two cold dense cloud
collide. Nevertheless, since cloud-cloud collisions may be considered
as the high-density version of the broader picture of colliding gas
flows, both pictures are consistent with a dynamic cloud and star
formation scenario. 


The remaining follow-up question relates to the origin of such
converging gas flows? Are the flows externally triggered, e.g., due to
nearby supernovae explosions or the spiral arm passage of the gas? Or
are the flows produced by the large-scale cloud collapse? The constant
flow rates from large to small spatial scales are consistent with a
self-similar, gravitationally driven cloud collapse. Furthermore, the
fact that the converging gas flow signatures are only seen in the
denser gas tracers, whereas the lower-density and more external CNM as
traced by HISA does not exhibit any significant velocity gradient, can
be considered as additional indication for the gas flow being
gravitationally driven in the dense gas region.  However, with the
given data, external compression as the cause of the converging flow
signatures cannot be excluded. It may even be that a combination of
external compression and gravitational collapse could explain the
data: The external driving by a spiral arm shock or supernova
compression may be the starting point of the collapse, which then may
rapidly be taken over by gravity. In that picture, externally
initiated converging gas flows and global gravitational collapse could
be part of the same overall scenario.


\section{Conclusions and Summary}
\label{conclusion}

The analysis of the kinematic parameters of the cloud surrounding the
IRDC G28.3 reveals clear signatures of two gas flows that converge at
the position of the central IRDC. The mass flow rather appears almost
constant from large to small spatial scales.  The spectral and spatial
signatures of the HISA compared to the other tracers are consistent
with a kinematic decoupling of the HISA-traced CNM from the denser
gas. Overall, the analysis is in general agreement with hierarchical
cloud structures that are dynamically evolving within converging gas
flows. The origin of such converging flows is consistent with a
self-similar, gravitationally driven collapse of the cloud. However,
converging gas flows could also be caused by external drivers, e.g.,
spiral arm shocks or supernovae driven shocks. The differentiation of
the possible different origins of the gas flows remains subject of
future investigations.

\begin{acknowledgements} 
  The National Radio Astronomy Observatory is a facility of the
  National Science Foundation operated under cooperative agreement by
  Associated Universities, Inc. HB likes to thank a lot for the
  inspiring discussions at the Harvard-Heidelberg Workshop on Star
  Formation (2019) about flow rates and their potential measurements,
  in particular to Diederik Kruijssen, Hope Chen, Shmuel Bialy and
  S\"umeyye Suri. HB, YW and JS acknowledge support from the European
  Research Council under the Horizon 2020 Framework Program via the
  ERC Consolidator Grant CSF-648505.  HB also acknowledges support
  from the Deutsche Forschungsgemeinschaft via SFB 881, “The Milky Way
  System” (sub-projects B1, B2 and B8).
\end{acknowledgements}

\bibliography{../../bibliography}   
\bibliographystyle{aa}    

\end{document}